\documentclass[a4paper]{article}
\RequirePackage[english]{babel}
\RequirePackage[latin1]{inputenc}
\RequirePackage[T1]{fontenc}
\RequirePackage{mathrsfs}
\RequirePackage{amsmath}
\RequirePackage{amssymb}
\RequirePackage{amsbsy}
\RequirePackage{bm}
\begin{document}
\title{\bf{On the geometry of the Dirac matter with\\ the Fermionic potentials and\\ its quantum properties}}
\author{Luca Fabbri\\ 
\footnotesize INFN, Sez. di Bologna \& Dipartimento di Fisica, Universit\`{a} di Bologna, 
Bologna, Italy}
\date{}
\maketitle
\begin{abstract}
We consider the torsional completion of gravity with electrodynamics for Dirac matter fields; we will see that these Dirac matter field equations will develop torsionally-induced non-linear interactions, which can be manipulated in order to be rearranged in the form of self-fermion potentials of a specific structure: eventually we will see that these non-linear interactions result into dynamical effects that are formally equivalent to those due to the quantum corrections.
\end{abstract}
\section*{Introduction}
In the present paper, we will consider the completely antisymmetric torsional completion of gravitation with electrodynamics as the underlying background hosting $\frac{1}{2}$-spin spinorial matter: that torsion is to be considered the completion of gravity comes from the fact that torsion and curvature can thus be thought as the strength of the potentials arising after gauging the space-time roto-translations, much in the same way in which the electrodynamic field is the strength of the potentials arising after gauging the phase \cite{h-h-k-n}, and the reason to insist on the complete antisymmetry of torsion is that in this case both principle of equivalence and causality can mathematically be implemented in an unambiguous way in order to interpret gravitation as contained within the curvature tensor \cite{h,xy,a-l,m-l,f/1a,f/1b}: the presence of torsion beside gravity, and beside also electrodynamics, has the consequence that then we can couple torsion to spin beside curvature to energy, and electrodynamics to the current, so that the matter field is coupled in the most extensive way \cite{ha}, and because torsion is taken to be completely antisymmetric the spin must be completely antisymmetric too, and the Dirac matter field results to be the only matter field that can be coupled consistently \cite{f/2a,f/2b}; one of most important consequences of the torsion-spin density coupling is that there arise torsionally-induced self-spinor interactions with interesting properties, like the fact that they can dynamically entail the principle of exclusion for fermions \cite{f/3}. Quite recently, this specific coupling has been written in its most general form \cite{Baekler:2011jt,Fabbri:2011kq}; additionally, it was shown that such non-linear potentials can allow one to solve the problem of positive-defined energies for fermions in a rather simple way \cite{Fabbri:2012ag}. Here, we will continue to investigate such non-linear potentials showing that they will give rise to effects that parallel those usually ascribed to quantum corrections.
\section{Kinematical Symmetries}
We begin with a summary of the notation, for which we will consider the fundamental fields given by the metric tensors as $g_{\alpha\sigma}$ and $g^{\alpha\sigma}$ symmetric and inverse of one another with connection $\Gamma^{\alpha}_{\mu\nu}$ defining a covariant derivative $D_{\mu}$ for which we have the condition $Dg\!=\!\!0$ and with torsion $Q^{\alpha}_{\phantom{\alpha}\mu\nu} 
\!\!=\!\!\Gamma^{\alpha}_{[\mu\nu]}$ taken completely antisymmetric for a transposition in any of its couple of indices: as it has been demonstrated in \cite{f/1a,f/1b} the metric-compatibility condition and complete antisymmetry of torsion respectively encode the fact that the metric can be flattened and there exists a unique symmetric part of the connection that can be vanished in the same neighborhood of the very same coordinate system and
\begin{eqnarray}
&\Gamma^{\mu}_{\sigma\pi}
=\frac{1}{2}Q^{\mu}_{\phantom{\mu}\sigma\pi}
+\frac{1}{2}g^{\mu\rho}\left(\partial_{\pi}g_{\sigma\rho}
+\partial_{\sigma}g_{\pi\rho}-\partial_{\rho}g_{\sigma\pi}\right)
\label{connection}
\end{eqnarray}
in general, and then we can also define the curvature tensor
\begin{eqnarray}
&G^{\mu}_{\phantom{\mu}\rho\sigma\pi}=\partial_{\sigma}\Gamma^{\mu}_{\rho\pi}
-\partial_{\pi}\Gamma^{\mu}_{\rho\sigma}
+\Gamma^{\mu}_{\lambda\sigma}\Gamma^{\lambda}_{\rho\pi}
-\Gamma^{\mu}_{\lambda\pi}\Gamma^{\lambda}_{\rho\sigma}
\label{Riemann}
\end{eqnarray}
antisymmetric in the first and second couple of indices, so with one independent contraction $\!G^{\alpha}_{\phantom{\alpha}\rho\alpha\sigma}\!\!=\!G_{\rho\sigma}$ with $\!G_{\rho\sigma}g^{\rho\sigma}\!\!=\!G$ called Ricci tensor and scalar with
\begin{eqnarray}
&G^{\mu}_{\phantom{\mu}\rho\sigma\pi}\!=\!R^{\mu}_{\phantom{\mu}\rho\sigma\pi}
\!+\!\frac{1}{2}(\nabla_{\sigma}Q^{\mu}_{\phantom{\mu}\rho\pi}
-\nabla_{\pi}Q^{\mu}_{\phantom{\mu}\rho\sigma})
\!+\!\frac{1}{4}(Q^{\mu}_{\phantom{\mu}\lambda\sigma}Q^{\lambda}_{\phantom{\lambda}\rho\pi}
-Q^{\mu}_{\phantom{\mu}\lambda\pi}Q^{\lambda}_{\phantom{\lambda}\rho\sigma})
\label{decomposition}
\end{eqnarray}
given in terms of the Levi-Civita metric covariant derivative $\nabla_{\nu}$ and the Riemann metric curvature tensor $R^{\mu}_{\phantom{\mu}\rho\sigma\pi}$ whose contraction is correspondingly given according to $R^{\alpha}_{\phantom{\alpha}\rho\alpha\sigma}\!\!=\!R_{\rho\sigma}$ with $R_{\rho\sigma}g^{\rho\sigma}\!\!=\!R$ called Ricci metric curvature tensor and scalar as they are usually called; this geometrical setting defined in terms of Greek indices and covariant under the most general coordinate transformation can equivalently be translated into another formalism defined in terms of additional Latin indices for which we get the supplementary covariance under the special Lorentz transformation, where from the metric tensor we extract the pair of dual bases of tetrads $\xi^{a}_{\sigma}$ and $\xi_{a}^{\sigma}$ such that they verify orthonormality conditions $\xi_{a}^{\sigma}\xi_{b}^{\nu}g_{\sigma\nu}\!=\!\eta_{ab}$ and $\xi^{a}_{\sigma}\xi^{b}_{\nu}g^{\sigma\nu}\!=\!\eta^{ab}$ in terms of the Minkowskian matrices, while the spin-connection $\Gamma^{i}_{j\mu}$ defining the covariant derivative $D_{\mu}$ is such that it gives $D\xi\!=\!0$ and $D\eta\!=\!0$ and for a connection with two different types of indices one cannot define torsion: these conditions imply (\ref{connection}) is
\begin{eqnarray}
&\Gamma^{b}_{\phantom{b}j\mu}=
\xi^{\alpha}_{j}\xi_{\rho}^{b}\left(\Gamma^{\rho}_{\phantom{\rho}\alpha\mu}
+\xi_{\alpha}^{k}\partial_{\mu}\xi^{\rho}_{k}\right)
\label{spin-connection}
\end{eqnarray}
and it is antisymmetric in the two Lorentz Latin indices, with curvature
\begin{eqnarray}
&G^{a}_{\phantom{a}b\sigma\pi}
=\partial_{\sigma}\Gamma^{a}_{b\pi}-\partial_{\pi}\Gamma^{a}_{b\sigma}
+\Gamma^{a}_{j\sigma}\Gamma^{j}_{b\pi}-\Gamma^{a}_{j\pi}\Gamma^{j}_{b\sigma}
\label{Riemanngauge}
\end{eqnarray}
antisymmetric in both the coordinate and the Lorentz indices and writable according to the expression $G^{a}_{\phantom{a}b\sigma\pi}\!=\!G^{\mu}_{\phantom{\mu}\rho\sigma\pi} \xi^{\rho}_{b} \xi^{a}_{\mu}$ as expected. In an analogous way, we define the geometry of complex fields, where the introduction of the gauge-connection $A_{\mu}$ defines the gauge-covariant derivative $D_{\mu}$ that extends the differential properties to complex fields, and whose curvature is given by
\begin{eqnarray}
&F_{\mu\nu}=\partial_{\mu}A_{\nu}-\partial_{\nu}A_{\mu}
\label{Maxwellgauge}
\end{eqnarray}
antisymmetric in the two indices, called Maxwell tensor. In this formalism, the general coordinate transformation is translated into specific Lorentz transformation, whose explicit structure could be written in the usual real but also in a new complex representation, in which gauge fields are included, as we will show.

To write such a representation, we have to introduce the $\boldsymbol{\gamma}_{i}$ matrices belonging to the Clifford algebra from which we define $[\boldsymbol{\gamma}_{i}, \boldsymbol{\gamma}_{j}]\!=\!4\boldsymbol{\sigma}_{ij}$ such that they verify the condition $\{\boldsymbol{\gamma}_{i},\boldsymbol{\sigma}_{jk}\}\!=\!i\varepsilon_{ijkq} 
\boldsymbol{\pi}\boldsymbol{\gamma}^{q}$ implicitly defining the $\boldsymbol{\pi}$ matrix and where the $\boldsymbol{\sigma}_{ij}$ matrices are the infinitesimal generators of this representation, called spinorial representation, in terms of which spinors are defined, and it is possible to introduce the spinor-connection $\boldsymbol{A}_{\mu}$ through which we define the spinor-covariant derivative $\boldsymbol{D}_{\mu}$ containing the information about the dynamics of the spinor fields: the spinorial constancy of the matrices $\boldsymbol{\gamma}_{j}$ is implemented automatically, and thus the spinor-connection $\boldsymbol{A}_{\mu}$ is decomposed as
\begin{eqnarray}
&\boldsymbol{A}_{\mu}
\!=\!\frac{1}{2}\Gamma^{ab}_{\phantom{ab}\mu}\boldsymbol{\sigma}_{ab}\!+\!iqA_{\mu}
\label{spinor-connection}
\end{eqnarray}
in terms of the complex-valued spin-connection plus an abelian field which we may finally identify with the Maxwell gauge-connection, with curvature
\begin{eqnarray}
&\boldsymbol{F}_{\sigma\pi}
\!=\!\partial_{\sigma}\boldsymbol{A}_{\pi}\!-\!\partial_{\pi}\boldsymbol{A}_{\sigma}
\!+\![\boldsymbol{A}_{\sigma},\boldsymbol{A}_{\pi}]
\label{RiemannMaxwellgauge}
\end{eqnarray}
as a tensorial spinor antisymmetric in the tensorial indices writable as
\begin{eqnarray}
&\boldsymbol{F}_{\sigma\pi}
\!=\!\frac{1}{2}G^{ab}_{\phantom{ab}\sigma\pi}\boldsymbol{\sigma}_{ab}\!+\!iqF_{\sigma\pi}
\label{combination}
\end{eqnarray}
as a combination of Riemann and Maxwell tensors, in quite a compact way.

It is important for the following to define the discrete space and time transformations, where space reflections $\vec{x}\!\rightarrow\!-\vec{x}$ are such that they leave angles unmodified while flipping the rapidities which means that they swap left-handed and right-handed components in the transformation laws, and we have that similarly the time reversal $t\!\rightarrow\!-t$ is such that it leaves angles unmodified while flipping the rapidities and so again it swaps left-handed and right-handed components in the transformation laws: we assign the transformation for reflections as given by $\psi\!\rightarrow\!\boldsymbol{\gamma}^{0}\psi$ and for reversals as given by $\psi
\!\rightarrow\!\boldsymbol{\gamma}^{1} \boldsymbol{\gamma}^{2} \boldsymbol{\gamma}^{3}\psi$ but we have also to notice that because both discrete transformations have the same effect of switching the opposite helicities of the spinors then we might have assigned the two transformations the other way around. But the definition we have given can be proven to be correct, although it will only be possible to justify it after the introduction of the matter field equations, as we will do.
\section{Dynamical Couplings}
Now that we have that the geometric kinematic quantities have been defined, next point that needs to be settled is the definition of the material quantities, and the eventual requirement of a link between the geometric fields, on the one hand, and the material quantities, on the other hand, in order to implement the dynamics encoded by the action or equivalently by the field equations, and the most exhaustive way in which the coupling between geometry and matter is established is given when torsion is coupled to the spin like curvature is coupled to the energy, and the derivative of the gauge strength is coupled to the current, and the most general way in which this can be done is with the completely antisymmetric torsion-spin coupling field equations taken beside the non-symmetric curvature-energy coupling field equations given by the following
\begin{eqnarray}
&Q^{\rho\mu\nu}=-aS^{\rho\mu\nu}
\label{torsion-spin}\\
\nonumber
&\left(\frac{8\pi k}{a}\!-\!\frac{1}{2}\right)\!\left(\frac{1}{4}\delta^{\mu}_{\nu}Q^{2}
\!-\!\frac{1}{2}Q^{\mu\alpha\sigma}Q_{\nu\alpha\sigma}
\!+\!D_{\rho}Q^{\rho\mu}_{\phantom{\rho\mu}\nu}\right)
\!+\!\left(G^{\mu}_{\phantom{\mu}\nu}-\frac{1}{2}\delta^{\mu}_{\nu}G
-\lambda\delta^{\mu}_{\nu}\right)+\\
&+8\pi k\left(F^{\rho\mu}F_{\rho\nu}-\frac{1}{4}\delta^{\mu}_{\nu}F^{2}\right)
=8\pi kT^{\mu}_{\phantom{\mu}\nu}
\label{curvature-energy}
\end{eqnarray}
together with the field equations for the gauge strength-current coupling as
\begin{eqnarray}
&\frac{1}{2}F_{\alpha\mu}Q^{\alpha\mu\rho}+D_{\sigma}F^{\sigma\rho}=J^{\rho}
\label{gauge-current}
\end{eqnarray}
with completely antisymmetric spin $S^{\rho\mu\nu}$ and non-symmetric energy $T^{\mu\nu}$ and current $J^{\rho}$ verifying the set of conservation laws given by
\begin{eqnarray}
&D_{\rho}S^{\rho\mu\nu}+\frac{1}{2}\left(T^{\mu\nu}-T^{\nu\mu}\right)\equiv0
\label{conservationspin}\\
&D_{\mu}T^{\mu\nu}
+T_{\rho\beta}Q^{\rho\beta\nu}-S_{\mu\rho\beta}G^{\mu\rho\beta\nu}+J_{\rho}F^{\rho\nu}\equiv0
\label{conservationenergy}
\end{eqnarray}
and together with
\begin{eqnarray}
&D_{\rho}J^{\rho}=0
\label{conservationcurrent}
\end{eqnarray}
verified as matter fields satisfy matter field equations; the only matter field that can in general be considered is the Dirac matter field because this is the only field with a completely antisymmetric spin and non-symmetric energy
\begin{eqnarray}
&S^{\rho\mu\nu}=\frac{i\hbar}{4}
\overline{\psi}\{\boldsymbol{\gamma}^{\rho},\boldsymbol{\sigma}^{\mu\nu}\}\psi
\label{spin}\\
&T^{\mu}_{\phantom{\mu}\nu}=\frac{i\hbar}{2}
\left(\overline{\psi}\boldsymbol{\gamma}^{\mu}\boldsymbol{D}_{\nu}\psi
-\boldsymbol{D}_{\nu}\overline{\psi}\boldsymbol{\gamma}^{\mu}\psi\right)
\label{energy}
\end{eqnarray}
and also with the current
\begin{eqnarray}
&J^{\rho}=q\hbar\overline{\psi}\boldsymbol{\gamma}^{\rho}\psi
\label{current}
\end{eqnarray}
where the matter field equations are given by
\begin{eqnarray}
&i\hbar\boldsymbol{\gamma}^{\mu}\boldsymbol{D}_{\mu}\psi-m\psi=0
\label{field}
\end{eqnarray}
defined with no additional constraints, as shown in \cite{f/2a,f/2b}: then we have that the completely antisymmetric torsion-spin coupling field equations and non-symmetric curvature-energy coupling field equations are given by
\begin{eqnarray}
&Q^{\rho\mu\nu}=-a\frac{i\hbar}{4}
\overline{\psi}\{\boldsymbol{\gamma}^{\rho},\boldsymbol{\sigma}^{\mu\nu}\}\psi
\label{torsionspincoupling}\\
\nonumber
&\left(\frac{8\pi k}{a}\!-\!\frac{1}{2}\right)\!\left(\frac{1}{4}\delta^{\mu}_{\nu}Q^{2}
\!-\!\frac{1}{2}Q^{\mu\alpha\sigma}Q_{\nu\alpha\sigma}
\!+\!D_{\rho}Q^{\rho\mu}_{\phantom{\rho\mu}\nu}\right)
\!+\!\left(G^{\mu}_{\phantom{\mu}\nu}-\frac{1}{2}\delta^{\mu}_{\nu}G
-\lambda\delta^{\mu}_{\nu}\right)+\\
&+8\pi k\left(F^{\rho\mu}F_{\rho\nu}-\frac{1}{4}\delta^{\mu}_{\nu}F^{2}\right)
=8\pi k\frac{i\hbar}{2}
\left(\overline{\psi}\boldsymbol{\gamma}^{\mu}\boldsymbol{D}_{\nu}\psi
-\boldsymbol{D}_{\nu}\overline{\psi}\boldsymbol{\gamma}^{\mu}\psi\right)
\label{curvatureenergycoupling}
\end{eqnarray}
and for the gauge strength-current coupling field equations
\begin{eqnarray}
&\frac{1}{2}F_{\alpha\mu}Q^{\alpha\mu\rho}+D_{\sigma}F^{\sigma\rho}
=q\hbar\overline{\psi}\boldsymbol{\gamma}^{\rho}\psi
\label{gaugecurrentcoupling}
\end{eqnarray}
where the spinor field verifies the spinorial matter field equations
\begin{eqnarray}
&i\hbar\boldsymbol{\gamma}^{\mu}\boldsymbol{D}_{\mu}\psi-m\psi=0
\label{matterequations}
\end{eqnarray}
and this is the most general system of field equations \cite{Fabbri:2011kq}. We notice that because it is from conservation laws that these field equations have been obtained, then we may think at them as obtained after a process of integration and therefore the constants $\lambda$ and $m$ have to be thought as integration constants; the four different constants $a$, $k$, $q$ and $\hbar$ match the four independent field equations, also acknowledging that such field equations are consistent. We will be back to this.

Now that the matter field equation has been defined it is possible to see that the assignment of the space reflections $\psi\!\rightarrow\!\boldsymbol{\gamma}^{0}\psi$ together with the assignment of the time reversals $\psi\!\rightarrow\!\boldsymbol{\gamma}^{1} \boldsymbol{\gamma}^{2} \boldsymbol{\gamma}^{3}\psi$ is the correct one while the other way around would have been incorrect instead; in fact, in the matter field equations we have that in the differential term there is a compensation in the action of the transformations $\vec{x}\!\rightarrow\!-\vec{x}$ and $\psi\!\rightarrow\!\boldsymbol{\gamma}^{0}\psi$ as well as in the action of the transformations $t\!\rightarrow\!-t$ and $\psi\!\rightarrow\!\boldsymbol{\gamma}^{1} \boldsymbol{\gamma}^{2} \boldsymbol{\gamma}^{3}\psi$ therefore justifying why they are respectively the space reflection and time reversal: thus the assignment given above respects this situation precisely. So we may take this definition when checking the discrete transformations of the set of field equations.
\section{Physical Effects}
It is now possible to separate all torsional curvatures and derivatives into the correspondent torsionless curvatures and derivatives plus the torsional contributions that can be converted through the torsion-spin field equations into spinorial potentials in the gravitational field equations for the Ricci tensor as
\begin{eqnarray}
\nonumber
&\left(R_{\mu\nu}+\lambda g_{\mu\nu}\right)
+8\pi k\left(F^{\rho}_{\phantom{\rho}\mu}F_{\rho\nu}-\frac{1}{4}g_{\mu\nu}F^{2}\right)
=-4\pi km\overline{\psi}\psi g_{\mu\nu}+\\
&+8\pi k\frac{i\hbar}{4}\left(\overline{\psi}\boldsymbol{\gamma}_{\mu}\boldsymbol{\nabla}_{\nu}\psi
+\overline{\psi}\boldsymbol{\gamma}_{\nu}\boldsymbol{\nabla}_{\mu}\psi
-\boldsymbol{\nabla}_{\nu}\overline{\psi}\boldsymbol{\gamma}_{\mu}\psi
-\boldsymbol{\nabla}_{\mu}\overline{\psi}\boldsymbol{\gamma}_{\nu}\psi\right)
\label{gravity}
\end{eqnarray}
with electrodynamic field equations given by
\begin{eqnarray}
&\nabla_{\sigma}F^{\sigma\rho}=q\hbar\overline{\psi}\boldsymbol{\gamma}^{\rho}\psi
\label{electrodynamics}
\end{eqnarray}
while the matter field equations are given instead as
\begin{eqnarray}
&i\hbar\boldsymbol{\gamma}^{\mu}\boldsymbol{\nabla}_{\mu}\psi
+\frac{3a}{16}\hbar^{2}\overline{\psi}\boldsymbol{\gamma}_{\mu}\boldsymbol{\pi}\psi
\boldsymbol{\gamma}^{\mu}\boldsymbol{\pi}\psi-m\psi=0
\label{Dirac}
\end{eqnarray}
showing that the gravitational and electrodynamic field equations are identical to those we would have in the torsionless case but the matter field equations are those we would have if we were with no torsion but with self-interactions of the matter field with the Nambu-Jona--Lasinio potentials whose strength is controlled by the coupling constant $a$ which is left to be determined while the gravitational constant $k$ is the Newton constant, $q$ is the electric charge which can be either positive of negative, the cosmological constant $\lambda$ is empirically known to be positive, but as we will discuss also $m$ will be taken to be either positive or negative. The constant $a$ will be taken to be positive so that the non-linear potentials give rise to an effective repulsion forbidding superposition of two fermions, although no matter the value of $a$ the non-linear interaction always vanishes for the superposition of opposite-helicity fermions, thus entailing a dynamical form of the principle of exclusion \cite{f/3}; allowing not only the charge but also the mass to be either positive or negative permits us to consider the common definition of the matter/antimatter duality in its utmost generality, where not only the charge but also the mass term has to be inverted in passing from matter to antimatter, and despite this might indicate that negative masses are present, not only it can be proven that only positive masses are present, but also that positive energies are ensured \cite{Fabbri:2012ag}. The principle of exclusion as an effective repulsive interaction and the positivity of energy for both matter and antimatter fields are two physical requirements commonly treated in the usual interpretation with tools such as defining the fields to be Grassmann-valued.

We remind that about the issue of space reflections and time reversals we have come up with a reasonable implementation of such discrete transformations respectively as $\psi\!\rightarrow\!
\boldsymbol{\gamma}^{0}\psi$ and $\psi\!\rightarrow\!\boldsymbol{\gamma}^{1} \boldsymbol{\gamma}^{2} \boldsymbol{\gamma}^{3}\psi$ and which we are now going to employ in order to see that a peculiar fact occurs: when we use space reflections according to $\psi\!\rightarrow
\!\boldsymbol{\gamma}^{0}\psi$ the matter field equations are left invariant but when we use time reversals according to $\psi\!\rightarrow\!\boldsymbol{\gamma}^{1} \boldsymbol{\gamma}^{2} \boldsymbol{\gamma}^{3}\psi$ we flip the sign of the spin-contact term so that the matter field equations are not invariant any longer, which is quite a weird circumstance because it shows that it is possible to find a reasonable implementation of the time reversal discrete transformation for which the matter field equations are not symmetric: it is intriguing that the charge conjugation operation $\psi\!\rightarrow\!\boldsymbol{\gamma}^{2}\psi^{\ast}$ flips the non-linear term of the matter field equations but also the time reversal operation $\psi \!\rightarrow\!
\boldsymbol{\gamma}^{1} \boldsymbol{\gamma}^{2} \boldsymbol{\gamma}^{3}\psi$ flips the spin-contact term of the matter field equations while space reflections $\psi \!\rightarrow\! \boldsymbol{\gamma}^{0}\psi$ leave them invariant, so that the two spoiling effects of charge conjugation and time reversal cancel each other, and the total combined conjugation is a discrete transformation for which the matter field equations are symmetric, recovering a discrete symmetry that we know to be valid in quite general circumstances, as it is demonstrated by the L\"{u}ders theorem; although here we are under quite different assumptions, this parallel between matter/antimatter and time reversal transformation, and therefore the fact that under all discrete transformations combined together there is still symmetry, still holds. Nevertheless, we have to notice that, once again, one of the assumptions that here is missing is the presence of Grassmann variables, which are instead used in the commonly accepted approach.
\subsection{Interacting Matter and\\ Radiative Corrections}
In all what follows, since we are mainly interested in high-energy processes, we are going to neglect the gravitational field equations and all influence of gravity as gauge field into the remaining field equations, and so from now on we will focus solely on the flat-spacetime approximation of the electrodynamic field equations given as above according to the form
\begin{eqnarray}
&\nabla_{\sigma}F^{\sigma\rho}=q\hbar\overline{\psi}\boldsymbol{\gamma}^{\rho}\psi
\end{eqnarray}
and the matter field equations also given as before according to
\begin{eqnarray}
&i\hbar\boldsymbol{\gamma}^{\mu}\boldsymbol{\nabla}_{\mu}\psi
+\frac{3a}{16}\hbar^{2}\overline{\psi}\boldsymbol{\gamma}_{\mu}\boldsymbol{\pi}\psi
\boldsymbol{\gamma}^{\mu}\boldsymbol{\pi}\psi-m\psi=0
\end{eqnarray}
as the field equations we will employ. A further manipulation of the non-linear terms by means of some Fierz rearrangement will prove that such non-linearities can be written with a form of higher physical significance, as we shall see.

One of the things that we know from classical electrodynamics is that the field equations in the Lorentz gauge $\nabla_{\mu}A^{\mu}\!=\!0$ have general solution with the property for which the gauge potential is proportional to the current density given in terms of a proportionality factor containing the information about the form of the matter distribution, which is unknown to us in general, but we may look for special solutions, so that once this special solution is assigned it can be used to determine the correspondingly special form of the proportionality between current density and gauge potential: it is easy to see that
\begin{eqnarray}
&A_{\mu}=-\frac{2q\hbar^{2}}{3m^{2}b^{2}}\hbar\overline{\psi}\boldsymbol{\gamma}_{\mu}\psi
\label{gaugesolution}
\end{eqnarray}
is indeed solution of the gauge field equations, if we pick for a matter field distribution a special solution given according to the following expression
\begin{eqnarray}
&i\hbar\boldsymbol{\nabla}_{\mu}\psi\!-\!(1\!-\!b)P_{\mu}\psi
\!-\!\frac{3a}{16}\hbar^{2}\overline{\psi}\boldsymbol{\gamma}_{\mu}\psi\psi
\!-\!b\frac{m}{4}\boldsymbol{\gamma}_{\mu}\psi=0
\label{mattersolution}
\end{eqnarray}
with constraint given by $P_{\mu}\boldsymbol{\gamma}^{\mu}\psi\!=\!m\psi$ and for a constant $b$ generic; then it is easy to see that $F_{\mu\nu}\!=\!-\frac{4q\hbar}{3mb}
i\hbar\overline{\psi}\boldsymbol{\sigma}_{\mu\nu}\psi$ with $\nabla_{\rho}(\hbar\overline{\psi}\boldsymbol{\gamma}^{\rho}\psi)\!\equiv\!0$ as it is expected \cite{i}.

By employing the identities given by $2i\boldsymbol{\pi}\boldsymbol{\sigma}^{\mu\nu}\!=\!\varepsilon^{\mu\nu\alpha\beta}\boldsymbol{\sigma}_{\alpha\beta}$ one can show that the axial current is not conserved but it is given in terms of the relationship
\begin{eqnarray}
&\nabla_{\rho}(\hbar\overline{\psi}\boldsymbol{\gamma}^{\rho}\boldsymbol{\pi}\psi)
=2(m+\frac{3a}{8}\hbar^{2}\overline{\psi}\psi)(i\overline{\psi}\boldsymbol{\pi}\psi)
-\frac{27am^{2}b^{2}}{128q^{2}\hbar^{2}}F^{\mu\nu}F^{\alpha\beta}\varepsilon_{\mu\nu\alpha\beta}
\end{eqnarray}
for the partially conserved axial currents; then we have that
\begin{eqnarray}
&i\hbar\boldsymbol{\gamma}^{\mu}\boldsymbol{\nabla}_{\mu}\psi
-\frac{9amb}{32\hbar q^{2}}i\hbar qF^{\mu\nu}\boldsymbol{\sigma}_{\mu\nu}\psi
-(\frac{3a}{8}\hbar^{2}\overline{\psi}\psi+m)\psi=0
\end{eqnarray}
for the matter field equations. We can acknowledge that the partially conserved axial current has beside the correction to the mass term also the additional Adler-Bell-Jackiw type of correction while the matter field equations have beside the same correction to the mass term also the additional Pauli term.

If we want to consider only general matter field distributions we may not infer the precise form of the solution and therefore the exact form of the correction, but suppose for the sake of argumentation that it is somehow possible to get some sort of constraint, for instance due to some stable equilibrium condition, for which the relationship $am^{2}\!\approx\!q^{4}$ comes out as a link between the constants involved in the fermion-photon scattering: then we have the relationship
\begin{eqnarray}
&\nabla_{\rho}(\hbar\overline{\psi}\boldsymbol{\gamma}^{\rho}\boldsymbol{\pi}\psi)
=2(m\!+\!\frac{3a}{8}\hbar^{2}\overline{\psi}\psi)(i\overline{\psi}\boldsymbol{\pi}\psi)
-\left(\frac{27b^{2}}{128\hbar^{2}}\right)
q^{2}F^{\mu\nu}F^{\alpha\beta}\varepsilon_{\mu\nu\alpha\beta}
\end{eqnarray}
for the partially conserved axial current; also it is clear that
\begin{eqnarray}
&i\hbar\boldsymbol{\gamma}^{\mu}\boldsymbol{\nabla}_{\mu}\psi
-\left(\frac{9b}{16\hbar}\right)q^{2}\frac{iq\hbar}{2m}F^{\mu\nu}\boldsymbol{\sigma}_{\mu\nu}\psi
-(\frac{3a}{8}\hbar^{2}\overline{\psi}\psi+m)\psi=0
\end{eqnarray}
for the matter field equations. In all of them the corrections are those we would expect in general circumstances, apart from a fine-tuning of the $b$ parameter.

To interpret this result, we may say that in presence of non-linear terms the matter field must be given in terms of an extended matter distribution which can have interactions with the electromagnetic field it produces; analogously, the concept of radiative corrections means precisely that in the perturbative expansion higher-order terms, when they are not one-particle irreducible, describe a given particle in the process of absorbing the carrier of the interaction emitted by itself, which means that the particle must be a self-interacting object, and this is what non-linear terms for matter fields imply: the idea of radiative corrections explained by representing the electron as a rotating ring or disk or yet ellipsoidal shell is known \cite{Israel:1970kp,Lopez:1984hw} and here we have reported a way to look at the issue that is independent on the form of the matter distribution.

Despite that the torsion-spin coupling induces non-linear interactions that are incompatible with the linearity required in fermion field quantization, nevertheless these non-linear potentials and the field quantization appear to give rise to the same effects in terms of the anomalous terms in the partially conserved axial currents and effective interactions in matter field equations.
\subsection{Spinning Fermions and\\ Quantum Anomalies}
We have seen that charged extended fields have the possibility to interact with their own electrodynamic field, and now it would be interesting to see what happens if we forget about the charge of the field in order to consider only the spatial extension of the field as what gives rise to corrective terms, aiming to see what these corrections are in general: in order to do that it is essential first to investigate a way in which the conserved quantities of the theory given by the completely antisymmetric spin density tensor $S^{\rho\mu\nu}$ and the non-symmetric energy density tensor $T^{\mu}_{\phantom{\mu}\nu}$ with the current density vector $J^{\rho}$ can be rewritten in terms of other conserved quantities that will have a more immediate use for what we intend to do in the following: to this purpose, consider the plane-wave solution of the form $i\hbar\boldsymbol{D}_{\mu}\psi \!-\!P_{\mu}\psi\!=\!0$ for which the conserved quantities given by the spin and the energy densities (\ref{spin}-\ref{energy}) reduce to the form
\begin{eqnarray}
&S^{\rho\mu\nu}=\frac{i\hbar}{4}
\overline{\psi}\{\boldsymbol{\gamma}^{\rho},\boldsymbol{\sigma}^{\mu\nu}\}\psi\\
&T^{\mu\nu}=\overline{\psi}\boldsymbol{\gamma}^{\mu}\psi P^{\nu}
\end{eqnarray}
while the current (\ref{current}) is still
\begin{eqnarray}
&J^{\rho}=q\hbar\overline{\psi}\boldsymbol{\gamma}^{\rho}\psi
\end{eqnarray}
which resemble the conserved quantities of the macroscopic approximation we would have in the case of the Weyssenhoff spin fluid; because the Dirac field has a completely antisymmetric spin density we have that the spin density of the spin fluid must be taken in its completely antisymmetric form, while the energy density is expected to be symmetric, and they are given according to
\begin{eqnarray}
&S^{\rho\mu\nu}=\frac{1}{3}U^{[\rho}S^{\mu\nu]}\\
&T^{\mu\nu}=\overline{\psi}\psi m U^{\mu}U^{\nu}
\end{eqnarray}
while the current density has to be given by
\begin{eqnarray}
&J^{\rho}=\hbar\overline{\psi}\psi q U^{\rho}
\end{eqnarray}
which we have now to compare: employing the special Fierz identity given by the relationship $2\overline{\psi}\boldsymbol{\pi}\boldsymbol{\sigma}_{ik}\psi \overline{\psi}\boldsymbol{\gamma}^{i}\psi \!=\!
\overline{\psi}\boldsymbol{\gamma}_{k}\boldsymbol{\pi}\psi \overline{\psi}\psi$ we see that one must have
\begin{eqnarray}
&S^{\mu\nu}=\frac{3}{2}i\hbar\overline{\psi}\boldsymbol{\sigma}^{\mu\nu}\psi
\label{spindensity}\\
&\overline{\psi}\psi U^{\rho}=\overline{\psi}\boldsymbol{\gamma}^{\rho}\psi
\label{velocity}
\end{eqnarray}
where we have that $S^{\mu\nu}$ is the spin and $U^{\rho}$ the velocity of the particle; notice that while the velocity can be applied to point-like particles, the spin density necessarily needs to be applied to extended field distributions in order to be interpreted in a meaningful way. Extended matter field distributions have already been studied, especially for plane-waves in the macroscopic approximation \cite{Audretsch:1981xn}.

As we have done above, it is possible to consider the matter field equations manipulating them in order to get the validity of some sort of conservation laws, and it is possible to calculate the divergence of the axial current in which we may plug the spin density we have just obtained and where we may again introduce the modified mass according to $\frac{3a}{8}\hbar^{2}
\overline{\psi}\psi\!+\!m\!=\!M$ eventually getting
\begin{eqnarray}
&\nabla_{\rho}(\hbar\overline{\psi}\boldsymbol{\gamma}^{\rho}\boldsymbol{\pi}\psi)
=2Mi\overline{\psi}\boldsymbol{\pi}\psi
-\frac{a}{6}S^{\mu\nu}S^{\alpha\beta}\varepsilon_{\mu\nu\alpha\beta}
\end{eqnarray}
for the partially conserved axial currents; then we have
\begin{eqnarray}
&i\hbar\boldsymbol{\gamma}^{\mu}\boldsymbol{\nabla}_{\mu}\psi
+\frac{a}{4}i\hbar S^{\mu\nu}\boldsymbol{\sigma}_{\mu\nu}\psi-M\psi=0
\end{eqnarray}
as the matter field equations. Again we acknowledge that within the partially conserved axial current there is the Adler-Bell-Jackiw anomalous term while the matter field equations are those we would have with quantum anomalies.

Notice that in both the partially conserved axial currents and the matter field equations, it is the presence of torsion, which manifests itself as a spin-dependent term, what gives rise to the presence of anomalous terms, which means that in this context rotating-extended distributions of matter are what gives rise to quantum-like corrections; for quantum fields, partially conserved axial currents and matter field equations have anomalous terms and effective interactions that arise as a consequence of quantum corrections due to radiative processes, as deciphered by the form factor $F(q^{2})\boldsymbol{\sigma}^{\mu\nu}q_{\nu}$ which indicates that transversal degrees of freedom must be taken into account, and therefore rotating-extended fields have to be considered, as justified by the presence of centrifugal-repulsive interactions due to torsion: whether we consider the torsion-spin coupling as the centrifugal-repulsive barrier that gives rotating-extended matter distributions or interpreting the fact that such particles are not point-like because of the anomalous terms due to quantum corrections \cite{n-j--l}, there are similarities between torsionally-induced non-linear potentials and the field-quantization scheme.

As already mentioned, the torsionally-induced non-linear potentials and the fermion-field quantization appear to give rise to the same effects in terms of the anomalous terms in the partially conserved axial currents and effective interactions in matter field equations, displaying similarities that are so striking that it will certainly be worth studying a little further in forthcoming works.
\subsubsection{General Comments on the Parallel between\\ Rotating-Extended Distributions and\\ 
Spinor-Quantization Prescriptions}
In what we have done so far, we have considered the torsionally-induced non-linear potentials of the Dirac matter field equations trying to see what are the possible consequences for the dynamics of fermion fields; on the other hand, we paid constant attention to what happens in such a situation when it is treated in terms of the usual paradigm of field-quantization: we have seen that it is possible to assign a combined discrete transformation in such a way that the system of field equations is left invariant, and we have witnessed that all anomalous terms and effective interactions in partially conserved axial currents and matter field equations were obtained as well, whether we employ torsionally-induced non-linear potentials for fermionic Dirac matter fields or Grassmann-valued fermions with field-quantization yielding corrections. The situation for which torsionally-induced non-linear potentials and field-quantization corrections display such similarities is worth deepening; but beside the analogies it is also important to point out the differences. In the remaining section of the paper we will sketch a discussion about what discriminates these two approaches.

First of all, we have to point out the fact that while in the standard approach the Dirac matter field equation is symmetric under $C$, $T$ and $P$, separately, and thus jointly, in our approach the Dirac matter field equation is symmetric only under $P$, but not under $T$ and $C$ separately, although the joint $TC$, and therefore the full $PTC$, are still symmetries, as expected: this is due to the fact that according to our definition of time reversal and without Grassmann variables, the two discrete transformations $T$ and $C$ fail because of the presence of the non-linear term in the matter field equations; that this fact could address the problem of an arrow of time is something of which we have no understanding but if this circumstance were true then it would point toward the possibility that there would be no need for Grassmann variables. In absence of Grassmann-valued fields there would be no fermionic field-quantization, the entire construction relying upon it would fail, and it would thus be necessary to have quantum effects recovered in terms of something else; even better would be if those effects might be recovered precisely in terms of what renders impossible the employment of Grassmannian fields, that is the torsionally-induced non-linear potentials giving rise to self-interacting fermions. This is what we have been discussing.

In fact, in the exposed approach we have considered the anomalous terms and effective interactions in partially conserved axial currents and field equations, trying to justify their presence not in terms of field-quantization protocols but with torsionally-induced non-linear potentials. In the usual scheme of quantum field theory, such quantum anomalies are ascribed to radiative processes, given in terms of a perturbative expansion: the resulting, effective interaction has a general structure $F_{1}(q^{2})\boldsymbol{\gamma}^{\mu}\!+\!
F_{2}(q^{2})\boldsymbol{\sigma}^{\mu\nu}q_{\nu}$ and all information about the interaction is stored within $F_{1}$ and $F_{2}$ called form factors. The non-quantum order is obtained in the situation in which the form factor $F_{1}$ equals unity and the form factor $F_{2}$ vanishes, but quantum effects start to be present when radiative corrections render the form factor $F_{2}$ different from zero, and quantum anomalies are even more relevant when further radiative corrections are taken by considering higher-order terms: these form factors can be computed at any order, and computations taken order by order are extremely precise indeed, and this is what constitutes the success of quantum field theory; but if we were to ask what would be the entire perturbative expansion yielding the analytic expression of both form factors, then we would not know \cite{p-s}. In the present approach, the situation is different. The non-linearity of the theory makes impossible to have any approximation in the first place, so that it is not possible to obtain any result order by order, and in this respect our approach is certainly worse than the common approach in obtaining numerical values; so the result must be sought exactly, but at least in this respect our approach is not worse than the common one since neither of them is able to obtain the analytic expression in any way for the moment. Of course, this means that exact solutions will have to be found in order for the present approach to produce predictions as we intend to do in the immediate future of this research program.
\section*{Conclusion}
We have considered torsionally-induced non-linear interactions for the matter field equations showing that they could be rearranged in a form identical to that of the effective interactions due to quantum anomalies, and finding that the torsion-spin coupling for fermionic Dirac matter fields and the field-quantization scheme have a parallel worth deepening.

\end{document}